\documentclass[DIV16,twocolumn,10pt,headsepline,abstract=true]{scrartcl}
\usepackage[english]{babel}
\usepackage[utf8x]{inputenc}
\usepackage[intlimits]{amsmath}
\usepackage{graphicx}
\usepackage{float}
\usepackage{booktabs}
\usepackage{xspace}
\usepackage{url}
\usepackage{subfig}
\usepackage{color}
\usepackage[colorlinks,%
pagebackref=false, 
linktocpage=true, 
plainpages=false, 
bookmarksopen=true,%
bookmarksnumbered=true,%
pdfauthor={Pavel Hron},%
pdftitle={Numerical simulation of Escherichia
  coli growth in porous media},%
pdfsubject={},%
pdfkeywords={dycap, microbiological growth, monod
  kinetics, parameter estimation, experiments in heleshaw-cell},%
]{hyperref}        
\usepackage[expproduct=cdot,alsoload={hep,binary}]{siunitx}
\usepackage{nicefrac}

\addtokomafont{pagehead}{\small}
\addtokomafont{section}{\normalsize}
\addtokomafont{subsection}{\normalsize\normalfont}

\newunit{\years}{\text{year}}

\title{Numerical simulation of growth of \textit{Escherichia coli} in unsaturated porous media}
\author{P.~Hron$^1$, D.~Jost$^2$, P.~Bastian$^1$, C.~Gallert$^3$, J.~Winter$^2$, O.~Ippisch$^{1,4}$}
\date{}
\publishers{\small \begin{flushleft}\noindent $^1$University of Heidelberg, Interdisciplinary Center for Scientific Computing, Im Neuenheimer Feld 368, 69120 Heidelberg, Germany\\\vspace{2mm}
$^2$Karlsruhe Institute of Technologie, Institut f\"{u}r Ingenieurbiologie und Biotechnologie des Abwassers, Am Fasanengarten Geb. 50.31, 76131~Karlsruhe, Germany \\\vspace{2mm}
$^3$University of Applied Science Hochschule Emden/Leer, Faculty of Technology, Microbiology – Biotechnology, Constantiaplatz 4, 26723 Emden, Germany \\\vspace{2mm}
$^4$Clausthal University of Technology, Department of  Mathematics, Erzstra{\ss}e 1, 38678 Clausthal-Zellerfeld, Germany \\\vspace{2mm}
  Email: pavel.hron@iwr.uni-heidelberg.de
\end{flushleft}
}

\newcommand{\pH}{\mbox{$\textrm{pH}$}}
\newcommand{\dder}{\mbox{$\textrm{d}$}}
\newcommand{\Ecoli}{\textit{E.~coli}\xspace}
\newcommand{\DOC}{\mbox{$\mathrm{DOC}$\xspace}}
\newcommand{\Oxygen}{\mbox{$\mathrm{O_2}$\xspace}}
\newcommand{\Salpha}{\mbox{$s_{\alpha}\xspace$}}
\newcommand{\Sl}{\mbox{$s_l\xspace$}}

\newcommand{\Sg}{\mbox{$s_g\xspace$}}
\newcommand{\mumax}{\mbox{$\mu_{max,\cdot}\xspace$}}
\newcommand{\mumaxan}{\mbox{$\mu_{max,an}\xspace$}}
\newcommand{\mumaxa}{\mbox{$\mu_{max,a}\xspace$}}
\newcommand{\muan}{\mbox{$\mu_{an}\xspace$}}
\newcommand{\muanstar}{\mbox{$\mu_{an}^{*}\xspace$}}
\newcommand{\mua}{\mbox{$\mu_{a}\xspace$}}
\newcommand{\coxyw}{\mbox{$c_{l,O_2}\xspace$}}
\newcommand{\coxyg}{\mbox{$c_{g,O_2}\xspace$}}
\newcommand{\coxys}{\mbox{$c^*_{l,O_2}\xspace$}}

\newcommand{\cak}{\mbox{$c_{\alpha,\kappa}\xspace$}}

\newcommand{\kh}{\mbox{$k_H\xspace$}}
\newcommand{\khcc}{\mbox{$k_H^{cc}\xspace$}}
\newcommand{\X}{\mbox{$c_{l,X}\xspace$}}
\newcommand{\substrate}{\mbox{$c_{l,S}\xspace$}}
\newcommand{\subs}{\mbox{$s_i\xspace$}}
\newcommand{\ksi}{\mbox{$K_{si}\xspace$}}
\newcommand{\Ys}{\mbox{$Y_{S,\cdot}\xspace$}}
\newcommand{\Ysan}{\mbox{$Y_{S,an}\xspace$}}
\newcommand{\Ysa}{\mbox{$Y_{S,a}\xspace$}}
\newcommand{\Yo}{\mbox{$Y_{O_2}\xspace$}}

\newcommand{\Ko}{\mbox{$K_{O_2}\xspace$}}
\newcommand{\Ksan}{\mbox{$K_{S,an}\xspace$}}
\newcommand{\Ksa}{\mbox{$K_{S,a}\xspace$}}

\newcommand{\Bs}{\mbox{$B_{S,\cdot}\xspace$}}
\newcommand{\Bo}{\mbox{$B_{O_2}\xspace$}}

\newcommand{\thetal}{\mbox{$\theta_l\xspace$}}

\newcommand{\thetaalpha}{\mbox{$\theta_\alpha \xspace$}}

\newcommand{\Cells}{\mbox{$\text{cells}\xspace$}}
\newcommand{\volumel}{\mbox{$V_l\xspace$}}
\newcommand{\volumeg}{\mbox{$V_g\xspace$}}
\newcommand{\kla}{\mbox{$k_L\alpha\xspace$}}
\newcommand{\rd}{\mbox{$r_d\xspace$}}
\newcommand{\rp}{\mbox{$p_d\xspace$}} 
\newcommand{\mo}{\mbox{$m_o\xspace$}}

\newcommand{\agw}{\mbox{$a_{gw}\xspace$}}

\DeclareSIUnit\ecoli{\Ecoli{}}
\DeclareSIUnit\doc{\DOC}
\DeclareSIUnit\consumable{consumable}
\DeclareSIUnit\dw{dw}
\DeclareSIUnit\biomass{biomass}
\DeclareSIUnit\biomass{water}
\DeclareSIUnit\consumed{consumed}
\DeclareSIUnit\substr{substrate}
\DeclareSIUnit\glucose{glucose}
\DeclareSIUnit\oxygen{\Oxygen}
\DeclareSIUnit\cells{\Cells{}}
\DeclareSIUnit\pvolume{CF~volume}
\DeclareSIUnit\pper{per}

\newcommand{\cdm}{\litre}
\newcommand{\ccm}{\milli \litre}
\newcommand{\temp}{$21\pm 1$\si{\celsius}}

\newcommand\T{\rule{0pt}{2.6ex}}       
\newcommand\B{\rule[-1.2ex]{0pt}{0pt}} 

\begin{document}

\maketitle

\begin{abstract}\noindent {
A model for the aerobic and anaerobic growth of \textit{Escherichia
coli} (HB101 K12 pGLO) depending on the concentration of oxygen and
\DOC{} as substrate has been developed based on laboratory batch
experiments. Using inverse modelling to obtain optimal sets of
parameters, it could be shown that a model based on a modified double
Contois kinetic can predict cell densities, organic carbon
utilisation, oxygen transfer and utilisation rates for a large number
of experiments under aerobic and anaerobic conditions with a single
unique set of parameters.

The model was extended to describe growth of \Ecoli{} in unsaturated
porous media, combining diffusion, phase exchange and microbiological
growth. Experiments in a Hele-Shaw cell, filled with quartz sand, were
conducted to study bacterial growth in the capillary fringe above a
saturated porous medium. Cell density profiles in the Hele-Shaw cell
were predicted with the growth model and the parameters from the batch
experiments without any further calibration. They showed a very good
qualitative and quantitative agreement with cell densities determined
from samples taken from the Hele-Shaw cell by re-suspension and
subsequent counting. Thus it could be shown, that it is possible to
successfully transfer growth parameters from batch experiments to
porous media for both aerobic and anaerobic conditions.
}
\end{abstract}

\section{Introduction}

Bacterial growth and activity in the vadose zone is a topic of growing
interest for ecologists and soil scientists within the last decade
\cite{Cannavo2004, Holden2005, Or2007, Dechesne2008}. The knowledge
about microbial processes under unsaturated conditions was used
e.g. for in situ bioremediation methods \cite{Tindall2005, Lee2009} or
for sand filter techniques \cite{Bester2011}. However, there is still
the need for research on bacterial growth, i.e. in a complex region as
the capillary fringe \cite{Ronen1997}. We define the capillary fringe
(CF) broader than the classical definition only including the zone
where the gas phase is discontinuous down to the zero capillary
pressure isosurface, but as the region which is dominated by capillary
rise from the groundwater table.

Bacterial growth in soil usually depends on bioavailable water content
\cite{Skopp1990, Chang2003}, temperature \cite{Trevors1991, Bell2008}
and other physical or environmental parameters like pressure, pH
\cite{Fernandez-Calvino2011} or biogeochemical redox processes
\cite{Borch2010}. In addition to water saturation, the most important
factors controlling bacterial growth are the availability of substrate
or nutrients \cite{Reischke2013} and the availability of electron
acceptors like oxygen for aerobic or facultative anaerobic bacteria
\cite{Sierra1995} and nitrate for anaerobic respiration
\cite{Limmer1998}.

Normally most of the nutrients are transported in the soil via
percolating water after rainfall. The nutrient supply in the soil then
is highest at preferential flow paths of the water. Thus bacteria
induce ''hot spots of growth'' at these locations
\cite{Bundt2001}. Besides these preferential flow paths the CF offers
attractive growth conditions for aerobic soil microorganisms
\cite{Affek1998,Jost2010,Jost2011}. Water is sufficiently available
due to capillary rise and oxygen is delivered by diffusion in the
vadose zone. Nutrients or Nitrate as alternative electron acceptor are
replenished by percolating water.

The major aim of this paper is the description of bacterial growth in
the CF as a potential hot spot for the degradation of subsurface
contaminants either by aerobic processes in the unsaturated part of
the CF or by anaerobic processes in the fully saturated zone.

In some previous laboratory experiments \textit{Pseudomonas putida}, a
typical soil bacterium, was used to study biodegradation of toluene
within a contaminant-plume \cite{Bauer2008, Bauer2009} in a saturated
porous medium with a flow through chamber. In this study
\textit{Escherichia coli}, an indicator bacterium for faecal pollution
of the environment \cite{Chen2012}, was used. \Ecoli{} can grow under
anaerobic and aerobic conditions \cite{Clark1989, Madigan2010} which
both have to be considered in an appropriate growth model. Such a
growth model will be developed and parametrised based on batch
experiments, and applied to predict experiments in a flow through cell
with conditions representing the capillary fringe. The simulations are
performed without any additional calibration of the estimated
parameters.

\section{Batch experiments}
\subsection{Principal experimental setup}

To understand the growth of \Ecoli{} and to describe the population
dynamics (such as cell densities and their dependence on substrate
concentrations), a number of different batch-culture experiments under
varying conditions were performed. \Ecoli{} normally has its growth
optimum at a rather high temperature ($\approx$\SI{37}{\celsius})
\cite{Reiling1985,Lendenmann1995}, whereas for our experiments, the
cells were cultivated at room temperature \temp{} which is closer to
natural conditions.

\subsection{Bacterial strain and its cultivation}

For all experiments in this study the Gram-negative, motile bacterial
strain \Ecoli{} HB101 K12 pGLO (Bio-Rad, Germany) was used. The strain
was cultivated in autoclaved LB medium (with Tryptose~\num{10}, Yeast
extract~\num{5} and $\mathrm{NaCl}$ \SI{10}{\gram \per \cdm}) at a
starting \pH{} of \num{7.0}-\num{7.2}, and a dissolved organic carbon
(\DOC{}) concentration of \SI{6,8}{\gram \per \cdm}. The active cells
of this strain generate a green fluorescent protein (GFP) after
addition of \SI{3}{\mmol \per \cdm} Arabinose (sterile filtered, Roth,
Germany). Previous experiments showed that the \Ecoli{} strain could
not use L(+)-Arabinose for energy generation or biomass
production. Due to a linkage with the \textit{ampR-gen}, the strain is
also resistant against Ampicillin (\SI{0.3}{\mmol \per \cdm} of which
was added). While the fluorescence was not used in the work described
in this paper, the Arabinose was still added as it will be used in
future work.

\subsection{Aerobic growth of \Ecoli{}} \label{sec:aerobic-growth}

For the aerobic experiments cotton plugged shake flasks filled with
\SI{30}{\ccm} of the different solutions described below were used and
approximately \SI{2.3e7}{\cells \per \ccm} (taken from a freshly grown
culture) were added. Duplicate culture assays were shaken with a
frequency of \num{200}~rpm. The optical density (OD) at \SI{578}{\nm}
was measured after inoculation and after every \SI{1}{h} during the
initial phase (log-phase) of bacterial growth and at several points in
time afterwards with a photometer (M501 Single Beam Scanning
Spectrophotometer, Camspec, UK) to get the growth curves. Calibration
experiments showed that OD values were linear dependent on the cell
density.

The LB medium was used pure and diluted with \SI{0.9}{\percent}
$\mathrm{NaCl}$ solution to $\nicefrac{1}{2}$ (1:1), $\nicefrac{1}{5}$
(1:4) and $\nicefrac{1}{10}$ (1:9) of the original concentration. The
oxygen concentration in the air phase was kept constant at
\SI{20}{\percent}.

In addition the biomass (dry weight) of \Ecoli{} cells after the
growth phase was measured gravimetrically for each replicate. Cell
suspensions were centrifuged (\SI{8}{\minute}, \SI{12}{k}~rpm) and
washed once with autoclaved, deionised water after discarding the
supernatant medium. After a second centrifugation, the cells were
resuspended in deionised water, filled in ceramic crucibles and the
dry weight (\si{\dw}) of the biomass from each flask was determined.

The \DOC{} concentration of both, the original medium and the discarded
medium from the centrifuge, were measured with a ''High TOC'' carbon
analyser (Elementar, Hanau, Germany) and the amount of \DOC{} consumed
was determined.

Cell suspensions in undiluted LB medium reached their maximal OD of
approx. \num{4.4} after \SI{24}{\hour} of growth
(Fig.~\ref{fig:aerobic_contois}). \Ecoli{} cells in suspensions with a
lower \DOC{} concentration reached their highest OD later (after up to
\SI{32}{\hour}). The \Ecoli{} cells consumed only about
\SI{30}{\percent} of the available \DOC{}, which means a maximal
\DOC{} consumption of \SI{2.0}{\gram \per \cdm} for the cells grown in
undiluted LB medium.

The initial biomass (as average value of the dry weight of
the \Ecoli{} cells in two samples) was \SI{0.02}{\gram \dw \per \cdm},
the increase after \SI{24}{\hour} was \SI{1.69}{\gram \dw \per \cdm}
for growth in undiluted LB medium, \SI{0.9}{\gram \dw \per \cdm} in
1:1 diluted LB medium, \SI{0.42}{\gram \dw \per \cdm} in 1:4 and
\SI{0.27}{\gram \dw \per \cdm} in a 1:9 dilution.

\subsection{Anaerobic growth of \Ecoli{}} \label{sec:anaerobic-growth}

For the anaerobic experiments closed serum bottles with a total volume
of \SI{120}{\ccm}, each filled with \SI{20}{\ccm} of one of the four
different solutions and a nitrogen atmosphere (HiQ Nitrogen \num{5.0}
Linde Gas, Germany) were autoclaved. After an inoculation with
\SI{2.0e7}{\cells \per \ccm} the experiments were conducted under the
same conditions as for aerobic growth (only without continuous
shaking).

The amount of consumed or metabolised \DOC{} was estimated on the
basis of the hydrogen (according to \cite{Larsson1993}) and fatty acid
production. The metabolites of the mixed acid fermentation were
measured after \SI{48}{\hour} with a gas chromatograph (Varian 431-GC)
and a mass spectrometer (Varian 210-MS, Agilent Technologies,
USA). The OD was again measured at the start of the experiment, after
every hour in the log-phase and at some specific points afterwards,
and the biomass increase was measured for all samples at the end of
the experiment.

As expected, the growth of \Ecoli{} under anaerobic conditions was
slower than under aerobic conditions as a consequence of the lower
efficiency of substrate fermentation. A maximal OD of \num{0.45} was
reached after approximately \SI{24}{\hour} in undiluted LB
medium. About \SI{30}{\percent} of the available \DOC{} was consumed
or converted to metabolites. After \SI{24}{h} the biomass increase was
\SI{0.28}{\g \dw \per \cdm} for the cells grown in undiluted LB
medium.

\subsection{Aerobic growth of \Ecoli{} under different oxygen
concentrations} \label{sec:aerobic-growth-different}

To determine the dependence of the growth rate on oxygen availability,
experiments with different oxygen concentrations were conducted in
closed \SI{50}{\ccm} glass vessels filled with \SI{20}{\ccm} of
undiluted LB medium only. Before and after the addition of
\SI{2.3e7}{\cells \per \ccm}, each vessel was flushed (with a flow
rate of approximately \SI{1.5}{\cdm \per \hour}) with a mixture of
different ratios of sterile artificial air\footnote{\SI{20}{\percent}
oxygen and \SI{80}{\percent} nitrogen} and pure nitrogen\footnote{HiQ
Nitrogen 5.0 from Linde Gas, Germany} to achieve oxygen concentrations
of \num{0}, \num{1}, \num{2}, \num{3.5}, \num{4}, \num{6}, \num{10},
\num{15} and \SI{20}{\percent}. To achieve these concentrations,
dosing valves for low flow (Swagelok, Germany) were used for each
vessel. To guarantee optimal air input, the medium was stirred with
approx \num{200}~rpm on a magnetic stirring plate (Variomag,
USA). \SI{50}{\micro\liter} of \num{100} times diluted
silicon-antifoam emulsion (Carl Roth GmbH, Karlsruhe, Germany) was
added. Oxygen concentrations at regular time intervals were measured
by a non-invasive optode technique \cite{Haberer2011}.

The experiments showed that the growth rate of \Ecoli{} under aerobic
conditions was highly dependent on sufficient availability of
oxygen. At concentrations of \num{20}-\SI{21}{\percent \oxygen}
(corresponding to ordinary air) and \SI{6}{\percent}, the OD values
were almost identical. Below an oxygen concentration of
\SI{3.5}{\percent}, the growth rate decreased significantly. Despite
the continuous shaking the optode measurements showed a lower oxygen
content in the liquid phase than in equilibrium with the gas
phase. The consumption of oxygen in water thus was faster than oxygen
dissolution from air.

\subsection{Oxygen consumption}

To measure oxygen consumption and biomass production, experiments were
conducted in closed serum bottles filled with \SI{10}{\ccm} of liquid
and \SI{110}{\ccm} sterile artificial air to guarantee an initial
oxygen concentration of \SI{20}{\percent} or a total oxygen volume of
approx. \SI{22}{\ccm}, respectively.

Undiluted LB medium and two dilutions (1:4 and 1:9) were used, to
assess the influence of nutrient availability on oxygen
consumption. The bacterial inoculum was the same as in the previously
described experiments and the bottles were again shaken with
\num{200}~rpm. The oxygen concentration in the liquid phase was again
measured with a non-invasive optode technique. After 24~hours a sample
was taken from the head space of each bottle and the oxygen
concentration in the gas phase was measured with a gas chromatograph
(Chrompack, model CP 9001, Engstingen, Germany) and total biomass
production was determined gravimetrically.

After \SI{52}{\hour} the biomass of \Ecoli{} cells grown in 1:4
diluted LB was \SI{0.36}{\g \per \cdm} and \SI{0.23}{\g \per \cdm} in
1:9 diluted LB medium. The oxygen consumption for \Ecoli{} in
\SI{20}{\ccm} 1:4 diluted LB was \SI{19.7}{\milli \g \oxygen} and for
cells grown in 1:9 diluted LB a value of \SI{10.1}{\milli \g \oxygen}
was measured.

\subsection{Conclusions from the batch experiments}

The following conclusions on the main factors controlling growth of
\Ecoli{} have been obtained from the batch experiments:

\begin{itemize}
\item Cell growth is faster in presence of oxygen. This is a
consequence of the higher energy efficiency of aerobic respiration
compared to anaerobic fermentation.

\item Anaerobic growth takes place only if the amount of available
oxygen is very low. If enough oxygen is available only aerobic
respiration is active.

\item Oxygen is still consumed after the growth phase (log-phase),
when all convertible \DOC{} is already depleted. The necessary
nutrients are most probably taken from intracellular resources.

\item At high cell and \DOC{} concentrations the dissolution of oxygen
in water is slower than oxygen consumption.

\item In all batch experiments only up to \SI{30}{\percent} of the
total \DOC{} was consumed or converted for both the aerobic and the
anaerobic growth.
\end{itemize}

\section{Modelling microbial growth}
\label{sec:kinet-models-micr}

Based on the results of the batch-culture experiments models for the
prediction of the growth of \Ecoli{} were derived depending on oxygen
and \DOC{} availability based on common approaches in the literature
and their performance was compared for reproduction of all
experimental data with a single set of parameters.

\subsection{Substrate-limited kinetics}

Various mathematical models have been proposed to quantitatively
describe microbial growth kinetics, see \cite{Monod1949, Moser1958,
Contois1959, Powell1967, Dabes1973, Koch1982}.

The biomass concentration under anaerobic conditions is described by a
first order differential equation relating the change of the biomass
concentration over time to the current biomass concentration \X{}
multiplied with a specific growth rate $\muanstar$
\begin{equation*} \frac{\dder \X}{\dder t} = \muanstar
\X.
\end{equation*} $\muanstar$ is the product of a maximal specific
growth rate $\mumaxan$ and a relative specific growth rate. $\mumaxan$
is a characteristic of all organisms and it is related to their
ability to reproduce. It is simply defined as the increase of biomass
per unit of time under optimal conditions (no limiting nutrients).

Common growth kinetics expressing the relative speed of growth
depending on the concentration \subs{} of a single substrate are given
by:
\begin{align} &\text{Monod} \quad
\frac{\subs}{\ksi+\subs}, \label{monodsk} \\ &\text{Moser} \quad
\left(1+\ksi \subs^{-\lambda_i}\right)^{-1}, \label{mosersk} \\
&\text{Tessier} \quad
1-\exp\left(-\frac{\subs}{\ksi}\right), \label{tessiersk} \\
&\text{Contois} \quad \frac{\subs}{B_i \X + \subs}. \label{contoissk}
\end{align} The substrate affinity constant (half-saturation constant)
$\ksi$ can be interpreted as a reflection of the affinity of the
bacterial cell towards the substrate $\subs{}$. It represents the
substrate concentration at which growth with half the maximal speed
occurs. Moser's constant $\lambda_i$ and the constant $B_i$ in the
Contois model do not have direct biological meaning. In our case
$\text{\subs{}} = \substrate$, which denotes the concentration of
bioconvertible nutrients and is only a part of the total $\DOC$ in the
medium.

Together with a Monod kinetic \eqref{monodsk} we obtain for example
\begin{equation} \muanstar = \mumaxan \frac{\substrate}{\Ksan +
\substrate}. \label{Eq:specific-growth}
\end{equation}

\subsection{Multiple-substrate kinetics}

While under anaerobic conditions the growth of \Ecoli{} or other
facultative anaerobic microorganisms depend only on the bioavailable
organic carbon concentration, under aerobic conditions the oxygen
concentration $\coxyw$ in the liquid phase has to be taken into
account as well. \cite{Kornaros1997} used a double Monod model to
describe this double nutrient limitation for the growth of
\textit{Pseudomonas denitrificans}. It combines two Monod kinetics in
a multiplicative form. The (aerobic) specific growth rate with a
double Monod model is given by
\begin{equation} \mua = \mumaxa \frac{\substrate}{\Ksa + \substrate}
\frac{\coxyw}{\Ko + \coxyw}. \label{Eq:double-specific-growth}
\end{equation} In both growth models \eqref{Eq:specific-growth} and
\eqref{Eq:double-specific-growth}, the Monod kinetics can be
substituted with each of the other models of the relative specific
growth rate \eqref{mosersk}-\eqref{contoissk}.

\subsection{Combination of aerobic and anaerobic growth}

While aerobic growth of \Ecoli{} cells is already occurring at low
oxygen concentrations, anaerobic growth might still be important. As
aerobic respiration is much more efficient, we assume that only
aerobic growth occurs if $\mua$ is higher than the growth rate for
purely anaerobic growth under the same substrate limitation ($\mua
\geq \mu_{an}^*$). Under these assumptions the specific growth rate
function for anaerobic growth in presence of oxygen $\muan$ can be
defined as
\begin{equation} \muan =
\max(\mu_{an}^*-\mua,0), \label{Eq:anaerobic-growth-term-modified}
\end{equation}

The total growth rate $\mu=\mua + \muan$ is then a non-decreasing
function of \coxyw{} for constant \substrate{}, where only the ratio
between aerobic and anaerobic growth is changing.

\subsection{Mass balance equations}

We consider a constant volume \volumel{} of liquid phase (culture
medium) and gas phase \volumeg{}. As there is no injection or removal
of cells during the experiment, the balance equation for the cell
density $\X$ is given by
\begin{subequations}
 \begin{align} \frac{\dder \X}{\dder t} &= \left(\mua + \muan - r_d
\right) \X , \label{Eq:grm-growth-cells} \\ \intertext{ where $r_d$ is
the decay rate. The generic balance equation for the consumable
substrate \substrate{} has the form } \frac{\dder \substrate}{\dder t}
& = - \left( \frac{\mua} {\Ysa} + \frac{\muan }{\Ysan} \right)
\X, \label{Eq:grm-growth-substrate} \\ \intertext{where the yield
coefficients $\Ysa$ and $\Ysan$ are the link between growth rate and
substrate utilisation, a measure for the efficiency of the conversion
of a substrate into biomass.} \intertext{Oxygen transfer between
liquid and gas phase occurs at the interface between both phases. For
gases with a low solubility, like oxygen in water, the local
equilibrium concentration of oxygen in water \coxys{} can be described
by Henry's law} \coxys &= \kh p_{O_2}, \notag \\ \intertext{where~\kh{}
is the Henry constant and $p_{O_2}$ is the partial pressure of oxygen.
If we assume the validity of the ideal gas law, we can rewrite Henry's
law in terms of a molar density $\coxyg =
\frac{p_{O_2}}{RT}$. Introducing a modified Henry constant $\khcc =
\frac{\kh}{RT}$, we get the relation $\coxys = \khcc \cdot \coxyg$
\cite{Sander1999}. As the batch experiments have shown, the assumption
of equilibrium between gas phase and liquid phase is not valid for
oxygen exchange during rapid growth phases. Thus we introduce a
kinetic mass transfer model depending on the difference of the actual
and equilibrium concentration of oxygen in the liquid phase, a
gas-liquid mass transfer capacity coefficient $\kla$ depending on
gas/liquid interfacial area (which was constant for the batch
experiments) and the ratio of the phase volumes. The mass balance
equations for oxygen in the liquid and gas phase are then given by }
\frac{\dder \coxyw}{\dder t}& = - \left( \frac{\mua} {\Yo} + \mo
\coxyw\right) \X + \kla \left(\coxys - \coxyw \right)
, \label{Eq:grm-oxygen-liquid}\\ \frac{\dder \coxyg}{\dder t} & = -
\kla \left(\coxys - \coxyw \right)
\frac{\volumel}{\volumeg}, \label{Eq:grm-oxygen-gas}
 \end{align} \label{Eq:grm-model}
\end{subequations}

where $\Yo$ is the yield coefficient for oxygen and \mo{} is the
oxygen consumption factor for maintenance.

This system of ordinary differential equations \eqref{Eq:grm-model},
together with appropriate initial conditions, specifies a modified
growth rate model for \Ecoli{} including simultaneous aerobic and
anaerobic growth, cell decay and a kinetic description for oxygen
transfer. It can be customised with different models of the relative
specific growth rate.

\section{Parameter estimation}

The growth model \eqref{Eq:grm-model} developed contains a number of
different parameters like maximal growth rates, half saturation
constants, yield coefficients, maintenance and mass transfer
coefficient etc, which need to be determined using the data obtained
from the batch experiments.

Many different ways to determine the model's kinetic parameters have
been developed. However, some of them are predisposed to inaccuracy
and can be applied only under limited conditions. A linearisation with
a Lineweaver-Burk plot was used to determine kinetic parameters,
e.g. in \cite{Kornaros1997, Ataai1985, Sierra1995}, but this method is
rather unreliable as it can greatly increase small errors in the
measurements. In the last two decades, software based on non-linear
regression is used to determine growth parameters, e.g. in
\cite{Senn1994, Kovarova1996, Kovarova-Kovar1998}. In this approaches
a separate set of parameters is obtained for every single experiment
or even for different stages of an experiment. Usually the parameters
are averaged later.

In this work, we want to use a parameter estimation approach to obtain
model parameters, which describe the measured quantities (like optical
densities, biomass densities, \DOC{} concentrations and oxygen
concentrations in air and in water) for all experiments
simultaneously. It uses an inverse model together with the initial
value approach \cite{Richter1990}, \cite{Marsili-Libelli1992}, for
details see \ref{parameterestimation}.  The
Levenberg-Marquardt-Algorithm was used to solve the optimisation
problem with sensitivities derived by numerical differentiation. The
forward model solves the system of equations \eqref{Eq:grm-model}
using a Runge-Kutta-Fehlberg method for systems of ordinary
differential equations. The initial conditions are derived from the
batch experiments.

\subsection{Parameter estimation for the batch experiments}

To combine the deviations between measured and simulated data for
different measured quantities into a single objective function
\eqref{Eq:residuals}, a suitable weighting factor $w_{ij}$ has to be
chosen. In theory this should be the inverse of the measurement
error. We assume that the error for each measurement is $10\%$ of the
maximal value of the measured quantity in the specific experiment.

To assess the predictive power of the growth model with the different
models of the relative specific growth rate (Monod, Moser, Tessier and
Contois model, \eqref{monodsk}-\eqref{contoissk}) a separate set of
parameters was estimated each of the combinations and the residua are
compared.

\subsubsection{Anaerobic growth}

In the absence of oxygen the growth of \Ecoli{} is purely anaerobic
($\mu=\muan$). This reduces the unknowns to the maximum growth rate
\mumaxan{}, the anaerobic substrate yield factor \Ysan{} and the decay
rate \rd{}, which both are the same for all four models of the
specific growth rate \eqref{monodsk}-\eqref{contoissk}, and parameters
for the growth kinetics like half saturation constant or Contois
saturation constant.

The decay rate was determined from four long-run anaerobic batch
experiments (6 days), each using a different dilution of the LB
medium. The remaining growth parameters were estimated from 8
experiments using 4 different dilutions of the LB medium (i.e. two
repetitions for each dilution) at an incubation temperature of \temp{}
(with a total of 125 single measurements).

\subsubsection{Aerobic growth}

The two types of aerobic experiments emphasise different aspects. In
the open system experiments the concentration of oxygen in air was
kept constant with a fast air circulation (see section
\ref{sec:aerobic-growth-different}) and thus the mass balance equation
for oxygen in the gas phase \eqref{Eq:grm-oxygen-gas} can be omitted,
while the closed system experiments allow a quantification of the
oxygen consumption.

A total of 334 individual measurements from 35 aerobic experiments
were used to estimate the remaining 7 or 8 parameters (depending on
the model used for the relative growth rate) which are only relevant
in the presence of oxygen.

\subsection{Results and Discussion}
\subsubsection{Performance of the different growth kinetics}

The residua obtained with the different models of the relative
specific growth rate are given in Table~\ref{Tab:resestimation}. There
was a generally good agreement between measured and simulated cell
biomasses $R_X$ and the final substrate concentrations $R_S$ as dry
biomass and substrate concentration were measured at the start of
batch experiment and after the log-phase only.

The best overall agreement between simulated and measured values was
obtained with the Contois model of relative specific growth. This is
mostly due to a much better reproduction of the measurements of
optical density and oxygen concentration in the liquid phase. The
models based on a Monod, Moser or Tessier kinetic perform
significantly worse. The obtained growth curves show a sharp switch
between log-phase and stationary/death phase, which was not observed
in the batch experiments and does not occur with the Contois
kinetics. The difference is even more striking, as the Contois based
model has the same amount of parameters as the models using the Monod
and Tessier kinetic and one parameter less than the model with the
Moser kinetic. Thus only the model with the Contois kinetic
\eqref{contoissk} is used in the rest of the paper.

\begin{table*}[t]
  \centering
  \caption{Residua gained by an inverse modelling for different
    kinetic models}
  \smallskip
  \small
  \label{Tab:resestimation}
  \begin{center}
    \small
    \begin{tabular}{l l l l l l l l l l l }
      \toprule
      &\multicolumn{4}{c}{anaerobic growth}&
      \multicolumn{6}{c}{aerobic growth} \\ \cmidrule(r){2-11}
      &$R_S$	&$R_{OD}$	&$R_X$	&$R$ &$R_S$	&$R_{OD}$  &$R_{g,O_2}$	&$R_{l,O_2}$	&$R_X$	&$R$\\ \midrule
      Contois	&1.8	&5.2	&6.7	&13.8&8.6	&14.1	&8.6	&9.2	&15.6	&56.2\\
      Monod	&1.2	&12.3	&7.2	&20.8&7.8	&39.4	&13.8	&25.1	&15.8	&101.9\\
      Moser	&1.8	&9.0	&6.9	&17.7&8.4	&34.3	&11.8	&43.5	&18.0	&116.1\\
      Tessier	&1.4	&12.9	&7.2	&21.5&8.4	&47.4	&14.9
      &28.0	&15.2	&113.8\\
      \bottomrule
    \end{tabular}
  \end{center}
\end{table*}

\subsubsection{Estimated parameters}

The estimated growth parameters for the model with the Contois kinetic
are given in Table~\ref{Tab:growth_parameters}. For all parameters the
standard deviations are relatively small indicating that the model is
both appropriate and not over-parametrised. Higher standard deviations
for the Contois saturation constants confirm the difficulty described
already in \cite{Kovarova-Kovar1998}, where the authors mentioned that
the saturation constant could vary even during a single growth cycle.

\begin{figure}[t] \centering
 \centering
 \includegraphics[width=0.5\textwidth]{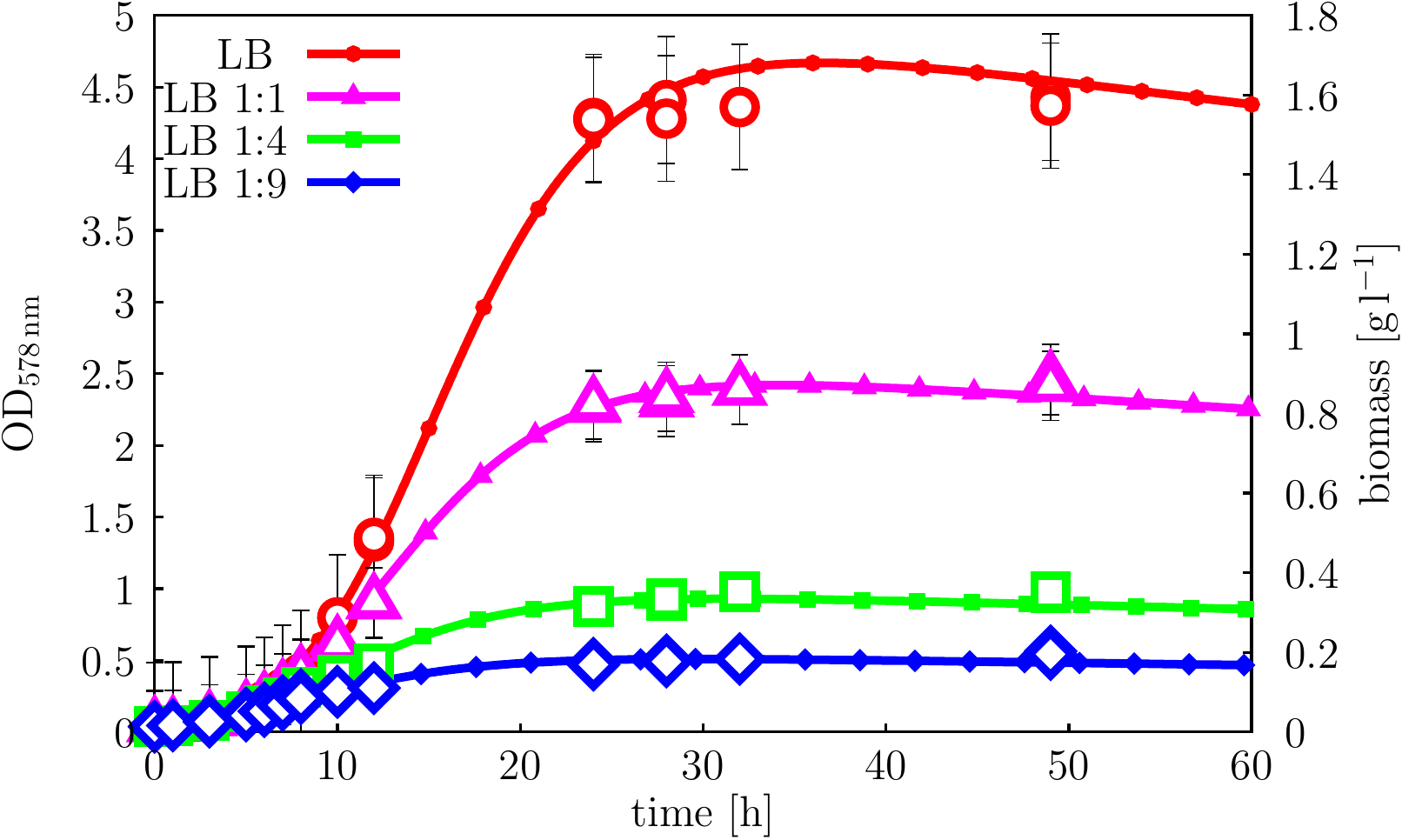}
 \caption{Comparison of OD for aerobic growth and
  growth curves as a solution of model \eqref{Eq:grm-model}
  with estimated parameters.}
 \label{fig:aerobic_contois}
\end{figure}

\begin{figure}[t] \centering
 \centering
 \includegraphics[width=0.5\textwidth]{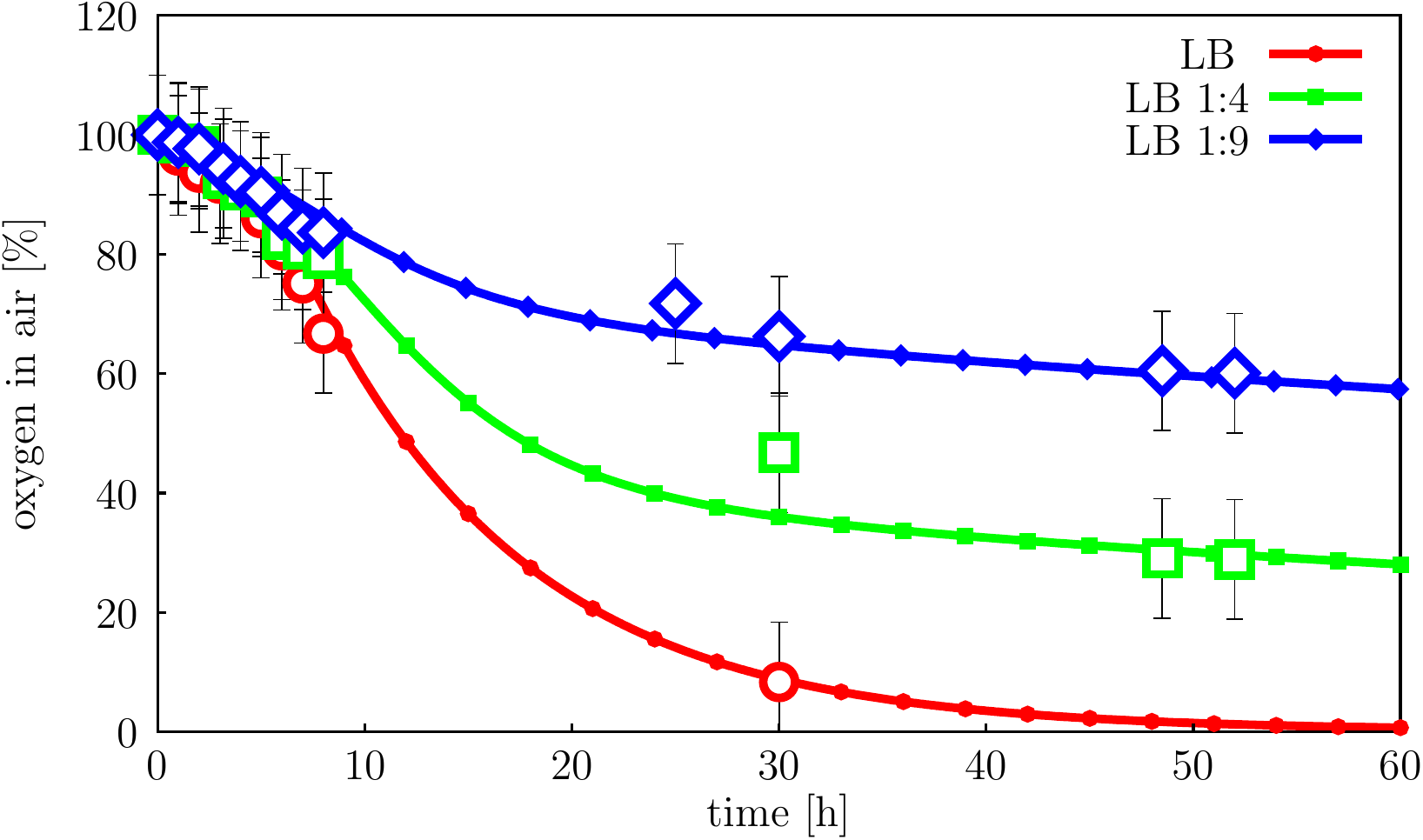}
 \caption{Comparison of measured data and oxygen concentrations in
  gas as a solution of model \eqref{Eq:grm-model}
  with estimated parameters.}
 \label{fig:oxygen_consumption_contois}
\end{figure}

However, the agreement between simulations and experiment was very
good for all types of measurements in all set-ups. As a typical
example the simulated and measured biomass concentration for aerobic
growth in closed serum bottles is shown in
Fig.~\ref{fig:aerobic_contois} for different substrate
dilutions. Fig.~\ref{fig:oxygen_consumption_contois} show the
corresponding decrease of the oxygen concentration in the gas phase.

The estimated maximal growth rates was $\SI{0.324}{\per\hour}$ for
aerobic growth and $\SI{0.255}{\per\hour}$ for anaerobic growth,
respectively. In \cite{Kovarova-Kovar1998} values for $\mumaxan$
between $0.19$ and $0.65$ in Chemostat and batch cultures at
$17-\SI{20}{\celsius}$ were listed. \cite{Reiling1985} and
\cite{Lendenmann1995} reported growth rates of \num{0.70} and
\SI{0.87}{\per \hour}, respectively for \Ecoli{} under optimal growth
conditions in a Chemostat culture.

Comparing these values with the results of our batch experiments for
aerobic growth of \Ecoli{}, it is obvious that the chosen growth
conditions (\temp{}) and LB medium were not optimal. Under natural
conditions the soil flora has to grow with a mixture of carbon
sources, mostly digestion products or amino acids draining from the
surface layer. The LB medium reflects naturally available carbon
sources relatively well. While the \DOC{} concentration in the LB
medium is very high compared to naturally oligotrophic conditions in
groundwater, an optimal nutrient supply was necessary in this study to
produce measurable quantities of biomass.

The yield coefficients of \SI{0.95}{\g \dw \pper \g \consumable \doc}
and \SI{0.49}{\g \dw \pper \g \oxygen} estimated for the \Ecoli{}
cells in LB medium are also in accordance with the yield coefficients
reported by \cite{Reiling1985}. Under anaerobic conditions bacteria
use a mixed acid fermentation (instead of respiration) to gain energy
for growth \cite{Stokes1949, Paege1961}, thus the growth is slower and
the yield coefficient decreases. According to our results, a yield of
\SI{0.163}{\g \dw \pper \g \consumable \doc} was determined, which is
close to the anaerobic yield coefficient of \SI{0.18}{\g \dw \pper \g
\glucose} which \cite{Ataai1985} found for some other strains of
\Ecoli{}.

These results clearly show that it is possible to describe the growth
of \Ecoli{} under a wide range of substrate and oxygen concentrations
including both aerobic and anaerobic growth with a single set of
realistic parameters.

\begin{table*}[t]
 \caption{Growth parameters derived by inverse modelling}
 \centering
  \smallskip
  \small\addtolength{\tabcolsep}{5pt}
 \label{Tab:growth_parameters}
 \begin{tabular}{ l l l }
\toprule
  Parameter& Aerobic& Anaerobic \T \B
  \\ \midrule
  Maximum growth rate \mumax{} [\si{\per\hour}] & $0.324\pm 3.7 \%$
  & $ 0.255\pm 6.9 \%$ \T \\
  Decay rate $r_d$ [\si{\per\hour}] & \multicolumn{2}{c}{ $\num{3.54e-3}\pm 4.2\%$ } \\
  Contois saturation c. \Bs{} [-]  & $1.81 \pm 15.3\%$& $3.07\pm 26.3 \%$ \\
  Yield for substrate \Ys{} [\si{\g \dw \per \g \consumable \doc}] &$0.95\pm 4.3\%$& $0.163\pm 2.1\%$ \\
  Contois saturation c. \Bo{} [-] & $ 0.019\pm 27.6 \%$ &-\\
  Yield for oxygen \Yo{} [\si{\g \dw \per \g \oxygen}] & $0.49\pm
  8.2 \%$ &-\\
  Maintenance for oxygen \mo{} [\si{\cdm \per \hour \per \g \dw}] & $\num{0.003}\pm
  19.9 \%$ &-\\
  Oxygen mass transfer c. \kla{} [\si{\per\hour}]&
  \multicolumn{2}{c}{$33.2\pm 9.3 \%$} \B \\ \bottomrule
 \end{tabular}
\end{table*}

\section{Growth of \Ecoli{} in the capillary fringe}

As this study is mainly targeted at the study of the growth of
\Ecoli{} in the capillary fringe, additional experiments with
Hele-Shaw cells were conducted to explore the microbial growth in a
porous medium. The results of these experiments were then used to test
the transferability of the results from the batch experiments (where
the micro-organisms were suspended in a liquid phase only) to a porous
medium without groundwater flow.

\subsection{Observation of bacterial growth in the capillary
fringe}

For the Hele-Shaw cells two $200\times 200 \times 3$~\si{\mm} glass
plates were fitted in silicone-greased spacers of \SI{2}{\mm}
thickness on each side and filled with quartz sand (diameter:
$0.35-0.71$\si{\mm}).  At the bottom of the glass cell, a stainless
steel reticule (\SI{25}{\mm} mesh size) was inserted as a carrier
layer for the sterile quartz sand.

At the beginning of the experiment the initially dry quartz sand in
the Hele-Shaw cells was vertically dipped \SI{1}{\cm} into a bacteria
suspension (approx \SI{2.4e7}{\cells \per \ccm}) with undiluted LB
medium. The chamber was kept fixed in this position for the next 3
days. A rapid rise of the water by capillary suction was observed
already in the first minutes. A (hydraulic) steady state was reached
after 6~hours \cite{Jost2011}. The top of the Hele-Shaw cell was open
to enable air circulation and continuous oxygen supply.

\subsection{Cell quantification in the CF sand volume}

As the suspended \Ecoli{} cells entered the porous medium with the
rising LB medium, the concentration was assumed to be homogeneous
initially and thus the cell density per volume soil resembles the
profile of volumetric water content.

Based on the results of the batch experiments all consumable DOC
should be depleted after 3 days either by anaerobic fermentation or by
aerobic respiration. Thus after this time span the Hele-Shaw cell was
lifted up from the suspension, disassembled and $1\times 1\times
0.2$~\si{\cubic \centi \metre} samples ($n=2$) of sand were taken at
positions with a vertical spacing of $\SI{1}{\cm}$ to count the
attached and suspended \Ecoli{} cells.

Each air dried \SI{0.2}{\cubic \centi \metre} sand volume was
transferred into \SI{0.9}{\percent} $\mathrm{NaCl}$ solution and was
shaken manually for several seconds \cite{Jost2014}. Then the
suspended cells were counted under the microscope according to
\cite{Taylor2002} and the cell density was converted into cells per
\si{\ccm} sample volume. The sample CF volume includes both the volume
of the matrix and of the pore space.

The rest of the porous medium was divided into \SI{1}{cm} high samples
to the gravimetric water content at different heights. Using a
porosity of 0.38 obtained from the dried samples, the gravimetric
water content was converted into volumetric water content. A van
Genuchten/Mualem model \cite{Genuchten1980} with $\alpha$ and $n$ as
free parameters and $m=1- \frac1n$ was fitted to the profile to
determine soil hydraulic properties.

\subsection{Cell densities in the CF}\label{sec:cell_density}

The profile of water saturation in the Hele-Shaw cell calculated with
the fitted van Genuchten model, and the vertical distribution of the
cell density of \Ecoli{} are shown in
Fig.~\ref{fig:ExperimentComparison}. The lower part of the domain up
to \SI{2}{\cm} is fully water saturated and the \Ecoli{} cells were
growing anaerobically. The cell concentration in this region was
\SI{2.2e8}{\cells \per \ccm \pvolume}. From the amount of
bioconvertible \DOC{}, the yield parameter \Ysan{} estimated from the
batch experiments and the cell densities, the average dry weight of
one \Ecoli{} cell was approx. \SI{5.e-13}{\gram}, which is in
agreement with values found in the literature.

Above the water table the water content decreases with increasing
height and the gas phase becomes continuous allowing for a much more
efficient oxygen transport. This allows aerobic growth and thus a
better substrate utilisation. The highest cell densities (up to
\SI{6.5e8}{\cells \per \ccm \pvolume}) were observed in a region with
a water content between $0.5$ and $0.8$, as here a optimal balance
between the availability of substrate and oxygen was obtained for the
\Ecoli{} cells. Above this zone the cell density decreases
proportional to the water content.

\subsection{Reactive transport model}

Microbial growth in an unsaturated porous medium can be simulated as a
special kind of reactive multiphase flow. For each component $\kappa$
a mass balance equation is given by
\begin{equation}
 \label{Eq:FullModelMoleBalance} \frac{\partial (\thetaalpha
\cak)}{\partial t} + \nabla\cdot\{j_{\alpha,\kappa} \} =
r_{\alpha,\kappa} + e_{\alpha,\kappa},
\end{equation}
where $\thetaalpha=\Salpha \phi$ is the volumetric
water or gas content depending on the phase saturation $\Salpha$ and
the porosity $\phi$, $\cak$ is the concentration of component $\kappa$
in phase $\alpha$, $j_{\alpha,\kappa}$ is the flux of component
$\kappa$ in phase $\alpha$, $r_{\alpha,\kappa}$ is the net change by
reactions and $e_{\alpha,\kappa}$ is the phase exchange. The main
components of the system are oxygen, consumable DOC as a dissolved
substrate and the microbial cell density (the two latter only occur in
the liquid phase).

Due to the high diffusion coefficient for oxygen in the gas phase and
the small solubility, we assume that pressure changes due to the
reactions are negligible and gas transport is purely diffusive. In the
absence of groundwater flow the solute is also transported by diffusion
only. We assume that \Ecoli{} cells are immotile and thus do not move
without water flow, therefore $j_{l,X}$ can be neglected.

Ficks' law is used together with the second model of Millington
and Quirk for the dependence of the effective diffusion coefficient on
phase saturation \cite{Jin1996} and one obtains
\begin{equation*} j_{\alpha,\kappa} = - s_{\alpha}^2 \phi^\frac{4}{3}
D_{\alpha,\kappa} \nabla \cak,
\end{equation*} where $D_{\alpha,\kappa}$ is the molecular diffusion
coefficient of component $\kappa$ in phase $\alpha$.

With the growth model \eqref{Eq:grm-model} derived from the batch
experiments we obtain the following reaction terms for cell density
$\X$ and substrate and oxygen concentration $\substrate, \coxyw$:
\begin{align*} r_{l,X}&=\thetal \left(\mua + \muan - r_d \right) \X, \\
r_{l,S}&= - \thetal \left( \frac{\mua}{\Ysa} + \frac{\muan }{\Ysan}
\right) \X, \\ r_{l,O_2} &= - \thetal \left( \frac{\mua} {\Yo} + \mo
\coxyw\right) \X.
\end{align*}

\subsection{Oxygen phase exchange in porous media}

To describe the non-equilibrium mass transfer of oxygen between gas
and liquid phase, we follow a model discussed in
\cite{Geistlinger2005} and \cite{Holocher2003} based on a stagnant
film model originally developed for spherical gas bubbles. A similar
approach was also used for the dissolution of a nonaqueous phase
liquid by \cite{Mayer1996}.

In this model it is assumed that the mass transfer rate
$e_{g,O_2}/e_{l,O_2}$ between both phases is proportional to the
difference between the concentration of oxygen in the liquid phase
$\coxyw$ and its equilibrium value $\coxys$, and the specific area of
the water-air interface $\agw$ and thus is given by
\begin{equation} - e_{g,O_2} = e_{l,O_2}=\beta \agw \left(\coxys -
\coxyw \right), \label{Eq:phaseExchangeModel}
\end{equation} where $\beta$ is a proportionality constant called mass
transfer coefficient.

In case of no flow, the mass transfer coefficient $\beta$ for a
spherical structure with a mean particle diameter $\rp$ is given by
\cite{Clift1978}
\begin{equation} \beta = \frac{1}{\delta_\mathrm{eff}} D_{l,O_2}=
\left(\frac{2}{\rp}+\frac{1}{\delta}\right)
D_{l,O_2},\label{Eq:masstransfercoeff}
\end{equation} where $D_{l,O_2}$ is the oxygen diffusion coefficient
in water and $\delta$ is the thickness of the stagnant film layer. In
no-flow conditions, the term $2 \rp^{-1}$ dominates and the effective
boundary layer thickness $\delta_\mathrm{eff}$ is therefore
proportional to $\rp$ \cite{Holocher2003}.

For this approach the determination of the gas-liquid interfacial area
is a crucial parameter. Many (empirical) relationships are proposed in
literature, for an overview see \cite{Porter2010}.

A rather simple model - based on geometrical considerations- to
estimate the total surface area for any packing of spheres was
presented by \cite{Gvirtzman1991}. They claim that the effective
interfacial area is proportional to the gas saturation and is given by
\begin{equation} \agw = \kappa \Sg
\frac{6\phi}{\rp}, \label{Eq:surfaceAreaBasic}
\end{equation} where $\kappa$ should describe the fraction of the
liquid-gas surface area exposed to mobile water. It is often assumed
to be equal to the porosity \cite{Geistlinger2005}.

A more complex approach to determine the gas-liquid interfacial area
using an equivalent pore size distribution derived from the water
retention function was proposed by \cite{Cary1994} and
\cite{Niemet2002}. The effective air-water interfacial area is the
given by
\begin{equation} \agw = \frac{3\phi}{2\sigma} \int_{0}^{\Sl}p_c(S)
\,dS, \label{Eq:surfaceAreaNiemet}
\end{equation} where $\sigma$ is the surface tension of the water. The
integral in \eqref{Eq:surfaceAreaNiemet} can be calculated
analytically for a van Genuchten parametrisation \cite{Genuchten1980}
with $m=1-\frac1n$ using the substitution $z=\frac{n+1}{n}$,
$w=\frac{n-2}{n}$ (see \cite{Niemet2002}, eq. (15)) and
\eqref{Eq:surfaceAreaNiemet} can be expressed as
\begin{equation} \agw = \frac{3\phi}{2\sigma} \frac{n-1}{\alpha n}
B(w,z)\left(1-I_u(w,z)\right), \label{Eq:surfaceAreaNiemetComputed}
\end{equation} where $B(w,z)$ is the beta function and $I_u(w,z)$
represents the incomplete beta function.

\subsubsection{Numerical discretisation}

The system of reaction-diffusion equations
\eqref{Eq:FullModelMoleBalance} was discretised using a cell-centered
finite-volume scheme in space and an implicit Euler scheme in
time. The arising non-linear equations are solved with
Newton-Method. The {DUNE/PDELab } software framework \cite{dune_2008,
pdelabalgoritmy} was used to facilitate the implementation.

As the flow process in the Hele-Shaw cell was essentially vertical, a
one-dimensional grid with a $1024$ cells was used, which produced a
grid convergent solution.

\subsection{Numerical simulation of growth in porous media}

As the initial infiltration and the establishment of a stationary CF
was fast (compared to the duration of the experiment) and no changes
in water saturation were observed afterwards, we can assume that the
consumable \DOC{} and \Ecoli{} concentrations at the beginning of the
experiment are uniform. The initial concentrations of all components
are given in Table~\ref{Tab:initial}.

All domain boundaries were assumed to be impermeable for components in
the liquid phase (zero-flux Neumann boundary condition). As the
Hele-Shaw cell was open at the top, a Dirichlet boundary condition was
used for oxygen in the gas phase with a mass fraction of $20.95\%$ and
standard pressure and temperature.

\begin{table}[h]
 \centering
 \caption{Initial concentrations of liquid and gas component}
 \smallskip
  \small\addtolength{\tabcolsep}{5pt}
 \label{Tab:initial}
 \begin{tabular}{l l }
  \hline
  Component & Initial concentration \\ \hline
   biomass $\X$ & \SI{12}{\milli \gram  \per \cdm} \\
  consumable substrate \substrate{} & \SI{2.0}{\gram \per \cdm} \\
  dissolved oxygen in water \coxyw{} & \SI{3.0}{\milli \gram \per \cdm} \\
  oxygen in air \coxyg{} & \SI{8.7}{\milli \mol \per \cdm} \\ \hline
  \end{tabular}
\end{table}

For the growth model the parameters determined in the batch
experiments were used (Table~\ref{Tab:growth_parameters}). As
described above the hydraulic parameters were obtained by independent
measurements (Section~\ref{sec:cell_density}) and the diffusion
coefficients were taken from literature (Table \ref{Tab:expcond}). No
additional calibration was performed.

\begin{table}[h]
 \centering
 \caption{Hydraulic parameters and diffusion coefficients
   for the sample calculations}
 \smallskip
  \small\addtolength{\tabcolsep}{5pt}
 \label{Tab:expcond}
 \begin{tabular}{l l }
  \hline
  \multicolumn{2}{l}{van Genuchten model} \T\B\\ \hline
   $n$ & 4.9 \T \\
   $\alpha$ & \SI{1.7e-3}{\per \pascal}
   \B \\ \hline
   porosity $\phi$& $0.38$ \T \\ \hline
   \multicolumn{2}{l}{Diffusion coefficients} \T\B \\ \hline
  $\dagger$$D_{l,S}$ & \SI{1.9e-10}{\meter \squared \per \second}
  \T \\
  $\ddagger$$D_{l,O_2}$ & \SI{2.2e-9}{\meter \squared \per \second}\\
  $\ddagger$$D_{g,O_2}$ & \SI{1.8e-5}{\meter \squared \per \second}\B\\ \hline
  \multicolumn{2}{l}{$\dagger$ \cite{Hendry2003}, $\ddagger$  \cite{Aachib2004}} \T \\
   \end{tabular}
\end{table}

\begin{figure}[h] \centering
  \includegraphics[width=0.5\textwidth]{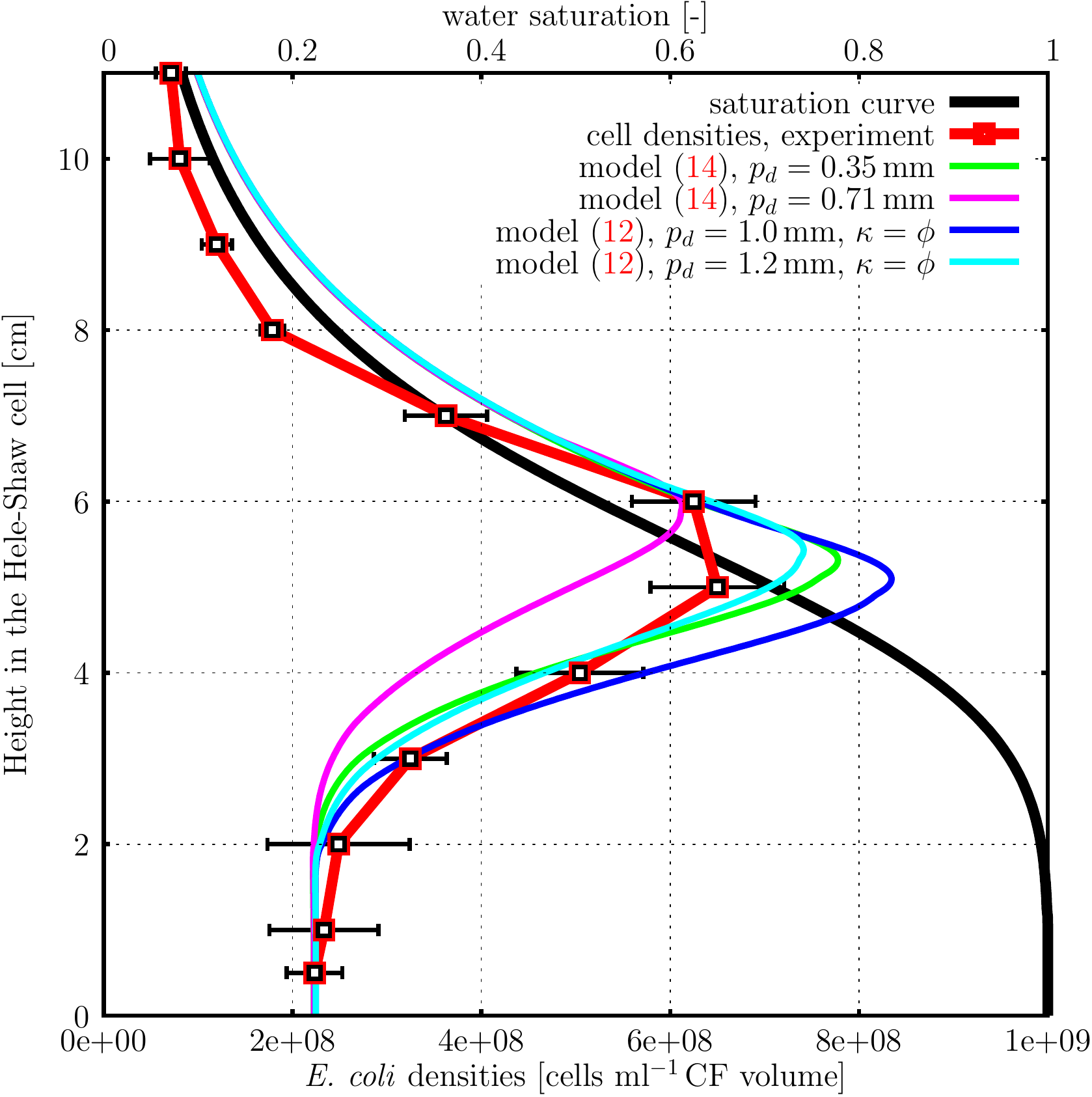}
 \caption{Comparison of cell density concentration computed by a
   numerical simulation for different oxygen exchange models with the
  experimental data after 3 days of growth in LB medium.}
 \label{fig:ExperimentComparison}
\end{figure}

A comparison of profiles of simulated and measured cell densities is
shown in Fig.~\ref{fig:ExperimentComparison}. The simulated cell
density is very sensitive to the model used to represent the phase
exchange of oxygen for water saturations between $0.5$ and $0.95$,
whereas in the completely water saturated part of the Hele-Shaw cell
growth is always anaerobic and at a water saturation below $0.5$
oxygen diffusion in the continuous gas phase and phase exchange are
always fast enough and thus growth is purely aerobic.

For both models of oxygen transfer, the efficiency of the phase
exchange decreases with increasing mean particle diameter $\rp$. Thus
for smaller values of $\rp$ the position of the peak of the cell
density is shifted to lower water saturations (higher positions) and
is at the same time decreasing in height due to the reduced amount of
substrate available at lower water saturations. While for the Niemet
model \eqref{Eq:surfaceAreaNiemetComputed} a good agreement between
measured and simulated values is obtained with $\rd= 0.35$~mm (which
is the size of the smaller grains in the sand used), a sufficient
match for the simpler model Gvirtzman and Roberts
\eqref{Eq:surfaceAreaBasic} requires an unrealistically large value of
$\rd=1.2$~mm.

In both models \eqref{Eq:surfaceAreaBasic} and
\eqref{Eq:surfaceAreaNiemetComputed} the gas-liquid interfacial area
is roughly proportional to the air content. However, the network
models \cite{Reeves1996, Held2001, Joekar-Niasar2007} and experiments
with glass beads \cite{Culligan2004, Porter2010} found this
relationship to hold only for $\Sl>0.3$. The air-water interfacial
area can decrease again for lower water saturations, which would
result in an overestimated phase exchange (and aerobic growth). This
could explain the slightly to high biomass in the region with low
water saturations. However, as oxygen is not the limiting factor under
these conditions this is rather unlikely.

\section{Conclusions}

\Ecoli{} in faeces naturally grows under anaerobic conditions by mixed
acid fermentation \cite{Madigan2010}, but it can also grow by
respiration e.g. in Chemostat cultures under aeration. Based on
experiments in batch culture, we developed a mathematical model, which
is able to consider both, the aerobic and the anaerobic growth of
\Ecoli{}.

A suitable growth model for \Ecoli{} (HB101 K12 pGLO) based on
laboratory experiments in batch culture was developed. The model is
able to consider both, aerobic and anaerobic growth
simultaneously. Using different types of data from many batch
experiments, a unique set of growth parameters describing all batch
experiments was determined using inverse modelling.

From a series of alternative growth rate models, the modified growth
rate model proposed by \cite{Contois1959} resulted in the best
agreement between measured and simulated data.

The proposed model for microbiological growth of \Ecoli{} in batch
cultures was extended with models for the oxygen phase exchange and
transport of substrate and oxygen to be able to describe growth in
porous media.

Results of the laboratory experiment in a Hele-Shaw cell filled with
quartz sand and of modelling of cell distribution showed a very good
prediction of the cell density profile of \Ecoli{} cells within the
CF. The numerical simulation was performed without any additionally
calibration of the parameters estimated from the batch
experiments. While the numerical simulation showed a strong dependency
on the model used to calculate air-water interfacial area, the model
of Niemet produced a very good agreement with the counted cell
densities using the particle diameter of the smaller grains in the
sand used in the experiment.

The very good predictive quality of the growth model developed from
batch experiments for the quite different conditions in a porous
medium makes it a promising basis for further experiments. While the
model was applied in this work only to bacterial growth under static
conditions, it will be applied to situations with groundwater flow and
continuous nutrient supply in the future.

\appendix
\section{Parameter Estimation}

\label{parameterestimation} The temporal development of a dynamic
system with the components $x_1,\dots,x_n$ is given by a system of
ordinary differential equations (in vector notation)
\begin{equation*} \frac{\dder \boldsymbol{x}}{\dder
t}=\boldsymbol{f}\left(\boldsymbol{x};\boldsymbol{\theta};t \right)
\quad \boldsymbol{x}(0)=\boldsymbol{x}_0
\end{equation*} with $\boldsymbol{x}=\left(x_1,\dots,x_n\right)^T$ and
parameters
$\boldsymbol{\theta}=\left(\theta_1,\dots,\theta_p\right)^T$. Let
$y_{i,j}$ denote measurements (or functions of the measurements) taken
at time points $t_j, j=1,\dots,m$.

The problem to be solved is the minimization of the objective function
(residuum)
\begin{equation} R(\boldsymbol{\theta})=\sum_{i=1}^n
R_i(\boldsymbol{\theta}) = \sum_{i=1}^n\sum_{j=1}^m
\left(\frac{x_{ij}(\boldsymbol{\theta},t)-y_{ij}
}{w_{ij}}\right)^2 \label{Eq:residuals}
\end{equation} over the set of admissible parameter values with
appropriate weighting factors $w_{ij}$.

{\small \noindent
\subsection*{Acknowledgements}
This study was funded by DFG (German Research Foundation) through
the Research Group FOR 831 ``Dynamic Capillary Fringes: A
Multidisciplinary Approach'' (Project Ga 546/5-2 and Ba
1498/7-2). Furthermore, we thank D.~Bonefas for his excellent
laboratory assistance.
}

{\small
\bibliographystyle{spmpsci}
\bibliography{references}
}
\end{document}